\documentclass[prb,preprint]{revtex4-1}

\usepackage{amsmath}  % needed for \tfrac, \bmatrix, etc.
\usepackage{amsfonts} % needed for bold Greek, Fraktur, and blackboard bold
\usepackage{graphicx} % needed for figures

\usepackage{amssymb}%
\usepackage{nicefrac}

\newcommand{\be}{\begin{equation}}
\newcommand{\ee}{\end{equation}}
\newcommand{\bea}{\begin{eqnarray}}
\newcommand{\eea}{\end{eqnarray}}

\newcommand{\bbm}{\begin{array}}
\newcommand{\ebm}{\end{array}}

\newcommand{\rme}{{\rm{e}}}

\begin{document}

\title{Hilbert's forgotten equation, the equivalence principle  and velocity dependence of free fall.}
% In a long title you can use \\ to force a line break at a certain location.

\author{David L.~Berkahn}
\affiliation{School of Electrical and Electronic Engineering, \\ University of Adelaide, SA 5005 \\ Australia}

\author{James M.~Chappell}
\email{james.chappell@adelaide.edu.au}
\affiliation{School of Electrical and Electronic Engineering, \\ University of Adelaide, SA 5005 \\ Australia}

\author{Derek Abbott}
\affiliation{School of Electrical and Electronic Engineering, \\ University of Adelaide, SA 5005 \\ Australia}

\date{\today}

\begin{abstract}
Referring to the behavior of accelerating objects in special relativity, and applying the principle of equivalence,
%, analogous to Einstein's original 1907 approach demonstrating the bending of light in a gravitational field, 
one expects that the coordinate acceleration of point masses under gravity will be velocity dependent. Then, using the Schwarzschild solution, we analyze the similar case of masses moving on timelike geodesics, which reproduces a little known result by Hilbert from 1917, describing this dependence. We find that the relativistic correction term for the acceleration based on general relativity differs by a factor of two from the simpler acceleration arguments in flat space.
As we might expect from the general theory, the velocity dependence can be removed by a suitable coordinate transformation, such as the Painlev{\'e}-Gullstrand coordinate system. The validity of this  approach is supported by previous authors who have demonstrated vacuum solutions to general relativity producing true flat space metrics for uniform gravitational fields. %This shows that the velocity dependence present in acceleration is not just consistent with general relativity, but is both a function of the coordinates and the particular metric solutions used. 
We suggest explicit experiments could be undertaken to test the property of velocity dependence.
\end{abstract}

\maketitle % title page is now complete

%\title{Hilbert's forgotten equation of velocity dependent acceleration in a weak gravitational field}
% Velocity dependence of accelerated systems, implications in gravitational fields, using the principle of equivalence---Hilbert’s lost equation.

%The principle of equivalence, the velocity dependence of accelerated systems and Hilbert’s lost equation 

%A simple use of the  principle of equivalence for free fall in gravity.

%\pacs{96.12.De,96.12.Bc}% PACS, the Physics and Astronomy
             
%%%% Subject entries to be placed here %%%%
%\subject{Physics}

%\keywords{Gravity, General relativity, Special
%relativity, Geodesics, Velocity
%dependence,Equivalence principle}%Use showkeys class option if %keyword

%\tableofcontents
%\begin{fmtext}
\section{Introduction}

General relativity provides the standard description of generalized motion
and gravity, 
%Geodesics of free fall particles follow space-time curvature in inhomogeneous gravitational fields. These are characterized by tidal %forces described by the geodesic deviation equations of a pseudo-Riemannian metric. 
where free fall particles follow geodesics within a given space-time metric. Inhomogeneous gravitational fields are characterized by tidal forces described by the geodesic deviation equations of a pseudo-Riemannian metric. 
The principle of equivalence although not playing a  highly prominent role in the modern version of the theory, is still useful if the limits of its applicability
 are clearly delineated. That is, the principle is believed to hold in the limit of small space-time regions of the gravitational field. We therefore begin by using the principle of equivalence for simple cases of acceleration from a special relativistic viewpoint, in an attempt to elucidate the qualitative result of velocity  dependence on acceleration for timelike geodesics in gravitational fields.   

Einstein reasoned that light would bend under gravity based on simple acceleration arguments in flat space and the principle of equivalence.  He then calculated a more precise result using general relativity, taking into account the effect of curved spacetime.  Using the same approach we deduce that acceleration under gravity is velocity dependent within accelerating frames under special relativity, which we then confirm with the Schwarzschild solution of general relativity---incidentally recovering a largely forgotten result by Hilbert. 

\subsection{Special relativity}

We define a spacetime coordinate differential with a four-vector 
\be
d x^{\mu} = ( c d t , d x , d y , d z ) ,
\ee
with contribution from three spatial dimensions and $ t $ is the time in a particular reference frame and $ c $ is the invariant speed of light~\cite{French1987}. In this paper we are able to focus exclusively on one-dimensional motion and so we can suppress two of the space dimensions writing a spacetime vector $ d x^{\mu} = ( c d t , d x ) $. We have the metric tensor $ g_{\mu \nu } = \left( \begin{array}{cc}
1 & 0 \\
0  & -1 \end{array} \right)$
that defines the covariant vector $ d x_{\mu} = g_{ \mu \nu } d x^{\nu} = ( c d t , -d x ) $. In the co-moving frame we have $ d x = 0 $ and so $ d x^{\mu} = ( c d \tau , 0 ) $, which defines $ \tau $ the local proper time.
We define the four-velocity 
\be
v^{\mu} = \frac{d x^{\mu}}{d \tau } = \frac{d t}{d \tau } \frac{d x^{\mu}}{d t }= ( \gamma c , \gamma v ) , 
\ee
where $ v = d x/dt $ and 
\be
\gamma = \frac{d t}{d \tau } = \frac{1}{\sqrt{1- \frac{v^2}{c^2} } }.
\ee
We then have the magnitude of the velocity four-vector
\be
\sqrt{v^{\mu} v_{\mu}}  = \sqrt{ \gamma^2 c^2 -\gamma^2 v^2 } = c 
\ee
that is a Lorentz invariant, where we have used the Einstein summation convention.  We also have the four-acceleration~\cite{Desloge1987} 
\be \label{properAccelerationDifferential}
a^{\mu}  = \frac{d^2 x^{\mu}}{d \tau^2 } = ( \gamma^4  a \frac{v}{c} , \gamma^4 a ) . 
\ee
Note that the special case of one-dimensional motion implies that $ v $ is parallel to $ a = \frac{d^2 x}{d t^2} $, which is not necessarily true in three-dimensions.  We then find the magnitude squared of the four-acceleration 
\be \label{magnitudeProperAcceleration}
a^{\mu} a_{\mu} =  \gamma^8 v^2 a^2/c^2 -\gamma^8 a^2  = -\gamma^6 a^2 .
\ee
Now, in the momentarily co-moving inertial frame (MCIF) we have $ v = 0 $ giving the acceleration  four-vector $ a^{\mu'} = ( 0 , \alpha ) $ and the four-velocity $ v_{\mu'} = ( c,0 ) $, which gives $ a^{\mu'} a_{\mu'} = -\alpha^2 $, as well as the expected orthogonality $ v^{\mu'} a_{\mu'} = 0 $.  Thus $ a^{\mu} a_{\mu} = -\alpha^2 $ is an invariant for all frames~\footnote{Because the four-acceleration is a four-vector, we can apply a Lorentz boost to the MCIF four-acceleration $ a^{\mu'} = ( 0 , \alpha ) $, with the transformation $ t = \gamma ( t' + v x'/c^2) $ and $ x = \gamma ( x' + v t') $. This produces $ a^{\mu} = ( \gamma v \alpha/c^2, \gamma \alpha ) $ and so comparing this with Eq.~(\ref{properAccelerationDifferential}) we have $ \gamma \alpha =  \gamma^4 a $ or $ \alpha =  \gamma^3 a $, confirming Eq.~(\ref{AlphaEquation}).}. Hence, comparing the magnitudes of the four-acceleration in Eq.~(\ref{magnitudeProperAcceleration}) with the magnitude in the MCIF we find $ \alpha = \gamma^3 a $. That is, in an alternate inertial frame we observe a coordinate acceleration~\cite{Rindler1991}
\be \label{AlphaEquation}
a = \alpha/\gamma^3 ,
\ee  
where $ \alpha $ is the acceleration observed in the MCIF.

\subsubsection{Thought experiment} \label{thoughtExperiment} 

Consider a rocket out in space far from the effects of any gravitational influences. Within this, effectively flat region of space, we place small frames of reference that individually can measure the acceleration of passing objects. We will call these types of frames PG1 for `particle group 1'.  The PG1 frames are currently at rest relative to the rocket and also with respect to each other and they are spread throughout the space surrounding the rocket. The rocket also has a hole at the top and bottom so that as the PG1 pass through they can measure the acceleration of the rocket. The rocket also has an inbuilt  mechanism so that, when the rocket is accelerating, it will drop a second group of particles, labeled PG2, from the top of the rocket, at predetermined fixed time intervals as measured by the rocket. Thus, each PG2 can also measure the rocket's acceleration, and at each instant when they are released will describe the MCIF for the rocket. Note, that as each PG2 are dropped at different times during the rocket's acceleration, then each will occupy a distinct inertial frame, whereas the PG1 all occupy the same inertial frame.

Now, for the sake of argument, let the rocket be accelerated at 9.8~ms$^{-2}$ and as specified, PG2 will be dropping from the top of the rocket. The rocket now accelerates with $ \alpha = T/m = 9.8~\rm{m s}^{-2} $ as measured by PG2, where $ m $ is the mass of the rocket and assuming $ T $ is an applied thrust in order to maintain a constant proper acceleration $ \alpha $.  As the rocket continues its acceleration it will encounter PG1 lying in its path that will enter the hole at the top of the rocket and while passing through measure the acceleration of the rocket.  Now, as the rocket is maintaining a steady acceleration, clearly the velocity of the rocket will be steadily increasing.  Hence the rocket will be encountering the PG1 frame at higher and higher relative velocities.

There are two questions we now wish to consider. Firstly:  {\it Will PG1 and PG2  measure the same acceleration for the rocket? }

Based on standard theory, we expect the answer to be in the negative. Intuitively, this will be because special relativity asserts that, as viewed by PG1, the rocket's velocity will converge to the light speed upper bound, and so the acceleration will appear to decrease. 
Since, this physical setting is described by Eq.~(\ref{AlphaEquation}), the one-dimensional relativistic equation for acceleration $ a $, as measured in the PG1 frames, can be written as 
\be \label{SRAcceleration}
a = \frac{\alpha}{\gamma^3}  = \frac{T}{m} \left ( 1 - \frac{v^2}{c^2} \right )^{3/2} ,
\ee 
where $ \alpha  $ is the acceleration measured in the co-moving inertial frames PG2 and $ v $ is the velocity of the rocket relative to any inertial frame measuring it, such as PG1. 
We note that the velocity dependence of  acceleration $ a $, viewed here by PG1, is clearly a frame dependent effect, as we can simply boost to the MCIF in order to remove it.
Of further interest is that due to the low acceleration of the rocket, we can ignore the effect of Einstein's time dilation occurring as a function of `vertical' position in the rocket, and hence the rocket can consider itself a satisfactory candidate for viewing the accelerations of PG1 and PG2. Observed accelerations are not necessarily reciprocal between frames, and we will show that when measured by the rocket the acceleration of PG1 is seen to be less than PG2.  

So, given the above, we can now ask a second question: {\it Given the principle of equivalence will these  results for accelerating observers, be replicated in a gravitational field? }

We presume that locally, in an approximately homogeneous section of a gravitational field, that the answer must be in the affirmative, provided we now replace the words `acceleration of the rocket with respect to both PG1 and PG2' with the words `fixed observer with respect to the source of gravitational field' and we replace the words `acceleration of PG1 and PG2 with respect to the rocket' with `PG1 and PG2 being freely falling particles on timelike geodesics'. In order to make this transition to gravitational fields, we now investigate the Schwarzschild solution of the general theory and the Rindler metric for accelerating frames.

\subsection{Gravitational fields}  

The central role played by the equivalence principle in the general theory was stated by Einstein in 1907:

{\it we [...] assume the complete physical equivalence of a gravitational field and a corresponding acceleration of the reference system. }

Einstein's equivalence principle (EEP) was initially based on the well established equivalence of gravitational and inertial mass, also called the weak equivalence principle, which has now been confirmed by experiment~\cite{Schlamminger2008} to an accuracy better than $ 1 \times 10^{-15} $.
It is now recognized that the full Einstein equivalence principle requires a curved spacetime metric theory of gravity in which particles follow geodesics, in accordance with the general theory~\cite{MisnerThorneWheeler1973}.

Incorporating the equivalence principle, our proposition is that since Eq.~(\ref{SRAcceleration}) pertains to a reference frame described above with an accelerating rocket, then this will also apply to a frame in an homogeneous region of a gravitational field.  That is, we write
\be \label{GravityAcceleration}
a = g \left ( 1 - \frac{v^2}{c^2} \right )^{3/2} ,
\ee 
where $ g $ represents the free fall of an object equivalent to PG2 but now in a gravitational field. Note that, since we are taking a single point in the rocket, we do not need to consider the full rocket frame\footnote{For a general accelerating frame depicted by the rocket we have the acceleration
\[ \label{ReviewerGeneralAccelerationIncluded}
a(X) = \frac{g (1 + g X/c^2 )}{[(g t/c)^2 + (1 + g X/c^2 )^2 ]^{3/2} },
\]
where $ X $ is the proper distance in the rocket, $ t $ is the coordinate time in the inertial frame and at  $ t = 0 $ the rocket is instantaneously at rest relative to the inertial frame.  Now, by specifying the weak field limit $ gX/c^2 \ll 1  $ we produce the  simpler
\[ \label{ReviewerGeneralAccelerationLowIncluded}
a(X) \approx \frac{g}{[(g t/c)^2 + 1 ]^{3/2} } = g (1 - v^2/c^2 )^{3/2} ,
\]
as assumed in Eq.(1.8), where we have used $ v = g t /\sqrt(1 + g^2 t^2/c^2) $.}.  
The proper acceleration that we have calculated refers to PG2, that is, the inertial frame instantaneously co-moving with the rocket. Note that the proper acceleration is the rate of change of proper velocity with respect to coordinate time, that is $ \alpha = \gamma^3 a = \frac{d \left (\gamma v \right )}{d t} = \frac{d}{dt} \left ( \frac{d x}{d \tau} \right )$. Hence to switch to the rocket frame we need to convert the coordinate time $ t$ to the proper time $ \tau $ of the rocket, and so we expect to pick up an extra factor of $ \gamma $, however, because clocks and rulers vary with position in the rocket, we need to set up a properly synchronized accelerating frame within which the acceleration is calculated, also called Rindler coordinates\footnote{
The Rindler metric for the rocket is given by
\[ \label{RindlerMetric}
c^2  d \tau^2 = \left (1 + \frac{g X}{c^2} \right )^2 c^2 d T^2 - d X^2 .
\]
}. This frame will produce the following acceleration from the perspective of the rocket frame, of
\begin{equation} \label{AccelerationRindlerAction}
 a = g \left ( 1 - \frac{2 v^2}{c^2} \right ) ,
 \end{equation}
where $ v $ is the velocity measured at the origin of rocket frame~\cite{Rowland2006}.  

%General relativity describe objects `stationary' in gravity as analogous to proper acceleration $ \alpha $. The point here is not to claim that this actually represents the true description in gravity in the sense of GR---clearly this is not a GR equation---the point is that the principle of equivalence used here tells us that we can expect some kind of dependence of the rate of free fall geodesics  on initial velocities. See appendix D for further clarification on this point.
%Hence, using an approach originally proposed by Einstein in his 1907 formulation of the equivalence principle, that is from a special relativistic view point, before the full spacetime curvature formulation, we produce a result 
%which predicts that a free-falling object (equivalent to PG1) in a homogeneous region of a gravitational field, is velocity dependent.This initial result now requires a more precise formulation in GR terms. Hence in the following section, we attempt to formulate a more accurate description of this by deriving a result using the Schwarzschild solution of general relativity.

 Hence, combining this result from special relativity with the principle of equivalence, we would predict that the acceleration of a free-falling object in a homogeneous region of a gravitational field, will be velocity dependent.
To support this conclusion, it has been shown, that constant acceleration  and a homogeneous uniform region of the gravitational field  can both be treated with a flat spacetime metric, consistent with general relativity and the Einstein vacuum field equations~\cite{munoz2010equivalence}.

We also note from
Eq.~(\ref{SRAcceleration})
that all inertial frames  will agree on \be \label{SRAccelerationInvariant1} \frac{a}{\left (1-\frac{v^2}{c^2} \right )^{3/2} } =\gamma^{3}(v)\frac{dv}{dt}  = \alpha .
\ee
Following Rindler~\cite{Rindler1991} and integrating Eq.~(\ref{SRAccelerationInvariant1}), where we choose $t=0$ when $v=0$ and $\alpha t=\gamma v$, we find 
\be \label{SRInvariant} \tau ^{2}=x^{2}-c^{2}t^{2} = \frac{c^{4}}{\alpha ^{2}} = \frac{c^4}{a^2} \left(1-\frac{v^2}{c^2}\right)^3 .
\ee
This shows, for inertial observers at least, a coordinate  invariant  view with respect to  constantly  accelerating  points indeed exists. 
%At this stage we simply note this result but continue with a coordinate dependent analysis. 

\section{Schwarzschild solution}

For a static, non-rotating, spherical mass the field equations of general relativity give the Schwarzschild solution
\be   \label{SchwarzschildMetric}
  c^2 d\tau^2  =  \left (1-\frac{2 \mu}{r} \right )c^2dt^2 - \left (1-\frac{2 \mu}{r} \right)^{-1} dr^2 
	 - r^2d\theta^2 -    r^2\cos^2\theta d\phi^2, 
\ee
where $ \mu = G M/c^2 $ and $ r $ is measured from the center and outside the mass~\cite{MisnerThorneWheeler1973}.
We also have the geodesic equation 
\be  \label{GeodesicEquation}
a^{\alpha} = \frac{d v^{\alpha} }{d \tau} = -\Gamma^{\alpha}_{\mu \nu } v^{\mu} v^{\nu} .
\ee
Now, from the metric we have $ g_{rr} = - \left (1-\frac{2 \mu}{r} \right)^{-1} $ and $ g_{tt} =  \left (1-\frac{2 \mu}{r} \right ) $ and so if we select purely radial motion, then we have the non-zero Christoffel symbols\footnote{The Christoffel symbol is defined as $\Gamma^{\alpha}_{\mu \nu} = \frac{1}{2} g^{\alpha \beta} \left ( \frac{\partial g_{\beta \mu}}{\partial x^{\nu}}  + \frac{\partial g_{\beta \nu}}{\partial x^{\mu}} - \frac{\partial g_{\mu \nu}}{\partial x^{\beta}}   \right )$.} 
\bea
 \Gamma_{rr}^r & = & \frac{1}{2} g^{rr} \partial_r g_{rr} = -\frac{\mu}{r^2} \left ( 1 - \frac{2 \mu}{r} \right )^{-1} = -\Gamma_{rt}^t  \\ \nonumber
\Gamma_{tt}^r & = & -\frac{1}{2} g^{rr} \partial_r g_{tt} = \frac{\mu}{r^2} \left ( 1 - \frac{2 \mu}{r} \right ) . \nonumber
\eea
The radial coordinate acceleration is then
\bea
\frac{d^2 r}{ d \tau^2 } & = & -  \Gamma_{rr}^r \left (\frac{d r}{d \tau} \right )^2 -\Gamma_{tt}^r \left (\frac{d t}{d \tau} \right )^2  \\ \nonumber
& = &  \frac{\mu }{r^2 }  \left ( 1 - \frac{2 \mu}{r} \right )^{-1}  \left (\frac{d r}{d \tau} \right )^2 -\frac{\mu c^2 }{r^2 }  \left ( 1 - \frac{2 \mu}{r} \right )  \left (\frac{d t}{d \tau} \right )^2  \\ \nonumber
& = &\frac{\mu c^2 }{r^2 }  \left (  \left ( 1 - \frac{2 \mu}{r} \right )^{-1}  \left (\frac{d r}{c d \tau}\right )^2 -\left ( 1 - \frac{2 \mu}{r} \right )  \left (\frac{d t}{d \tau} \right )^2  \right ) . \nonumber
\eea
%Applying the law that falling frames in gravity locally obey the rules of special relativity, We can consider the view of the falling frames and set  $\left(\frac{dr}{d\tau }\right)^2=v\left(1-\frac{v^2}{c^2}\right)^{^4}$.
The last term in brackets is simply the metric in Eq.~(\ref{SchwarzschildMetric}) and so equal to one, and so
\be \label{SecondDerivativeProperTimeMain}
a^r = \frac{d^2 r}{ d \tau^2 } = -\frac{\mu c^2 }{r^2 }  = -\frac{G M}{r^2 } .
\ee
For a rocket observer applying thrust in order to remain at a fixed radius $ r $ in the field, we have $ a^t =0 $ and so this implies the magnitude of the four-acceleration is
\be
\sqrt{g_{\mu \nu } a^{\mu } a^{\nu}} = \sqrt{g_{rr}} \frac{ G M}{r^2} = \frac{ 1}{\sqrt{1 - 2 \mu/r}} \frac{ G M}{r^2 } ,
\ee
which is a Lorentz invariant\footnote{For the observer at fixed spatial coordinates we have the four-velocity $ v^{\beta} = \left ( c \left ( 1- \frac{2 \mu}{r} \right )^{-1/2} , 0  \right ) $ and the four-acceleration $ a^{\beta} = \left ( 0, - \frac{ G M}{r^2}  \right ) $, where we have suppressed the two angular coordinates.  This observer then sees the four-velocity of the radial infalling particle $ u^{\beta} = \gamma \left (c \left ( 1- \frac{2 \mu}{r} \right )^{-1/2} , v \left ( 1- \frac{2 \mu}{r} \right )^{1/2}  \right ) $.}.

Now, we can write Eq.~(\ref{SecondDerivativeProperTimeMain}) as 
\be
\frac{d}{d \tau} \left ( \frac{1}{2} \left (\frac{d r}{ d \tau} \right )^2 - \frac{G M }{r} \right ) = 0 
\ee
and so
\be
 \frac{1}{c^2} \left (\frac{d r}{ d \tau} \right)^2 - \frac{2 G M }{c^2 r} = {\rm{constant}} = \frac{E_0^2}{m^2 c^4 } - 1,
\ee
where $ E_0 $ can be shown to be the conserved total energy of the particle.  Hence
\be \label{properTimeVelocityFormula}
\frac{d r}{ d \tau} = \pm c \sqrt{\frac{2 \mu }{r} + \frac{E_0^2}{m^2 c^4} -1 }.
\ee

Now $ \frac{d r}{d t} = \frac{d r}{d \tau} \frac{d \tau}{ d t} $ and so using Eq.~(\ref{SchwarzschildMetric}) we determine 
\be
\frac{dt}{d\tau} = \frac{E_0 }{m c^2 \left (1 - \frac{2 \mu }{r} \right ) }
\ee
and so we find
\be \label{VelocityCoordinate}
\frac{d r}{d t} =  \pm c  \left ( 1-\frac{2 \mu}{r} \right ) \sqrt{ 1 - \frac{m^2 c^4}{E_0^2} \left (1 - \frac{2 \mu }{r} \right )} .
\ee
Therefore, for particles entering the gravitational field with $ r $ approaching infinity with an initial velocity $ v $, we have $ E_0 = \frac{ m c^2}{\sqrt{1 - \frac{v^2}{c^2}}} $, as expected for flat space, where $ v = \frac{d r}{ d t} $.  For the alternate case of bound particles we have $ E_0 < m c^2 $, also calculated from Eq.~(\ref{VelocityCoordinate}).

Using the chain rule, $ \frac{d^2 r }{d t^2 } = \frac{d (dr/dt) }{d r } \frac{ d r }{ d t} $, we find
\be \label{SchwarzschildCoordinateAcceleration}
\frac{d^2 r}{d t^2} = - \frac{\mu c^2}{r^2 } \left ( 1 - \frac{2 \mu}{r} \right ) \left ( 3 \left ( 1 - \frac{2 \mu}{r} \right )\frac{m^2 c^4}{E_0^2} -2 \right ) .
\ee
Then, using the relation between $ E_0 $ and $ \frac{dr}{dt} $ we find
\be \label{SchwarzschildVelocityDependenceAppendix}
\frac{d^2 r}{d t^2} = - \frac{\mu c^2}{r^2 } \left ( 1 - \frac{ 2 \mu}{r} \right ) \left (1 - 3 \left (\frac{ d r}{c d t} \right )^2 \left ( 1 - \frac{ 2 \mu}{r} \right )^{-2} \right ) .
\ee

This shows the velocity dependence $ \frac{ d r}{d t} $ of the coordinate acceleration $ \frac{d^2 r}{d t^2} $ based on the Schwarzschild coordinates, for an observer at infinity. 
For the case of the weak field only we have $  \frac{2 \mu}{r} \rightarrow 0 $ and so
\be \label{SchwarzschildVelocityDependenceWeakField}
\frac{d^2 r}{d t^2} = - \frac{G M}{r^2 } \left (1 - \frac{3 v^2}{c^2} \right ) ,
\ee
a result first derived by Hilbert~\cite{hilbert1917grundlagen,Hilbert1924,mcgruder1982gravitational} in 1917, for the special case of particles moving radially in the Schwarzschild metric. We note that even in the weak field with very small distortion of space and time of the order $ \frac{2 \mu }{r} $ we can create a velocity dependence of order unity for relativistic particles.  We can perhaps see an obvious source for this with the velocity terms in the geodesic equation, in Eq.~(\ref{GeodesicEquation}). However, more fundamentally, we see that velocity dependent effects of a similar magnitude arise within special relativity in flat space, as shown by Eq.~(\ref{SRAcceleration}), showing that the velocity dependence effects arise on the basis of special relativity rather than gravitational spacetime distortions within the general theory.  Given this last point and that Eq.~(\ref{GravityAcceleration}) and  Eq.~(\ref{AccelerationRindlerAction}) not being reciprocal viewpoints as well as the true time asymmetry of accelerating frames in special relativity, we can assert that velocity dependent effects are not simply coordinate artifacts of general relativity.
The result of Hilbert also appears to indicate `gravitational
repulsion', for $ v > \frac{c}{\sqrt{3}}$, as indeed might also be claimed for Eq.~(\ref{AccelerationRindlerAction}). However we note that it applies to the case where the observer is at infinity, and as the coordinate velocity in Eq.~(\ref{VelocityCoordinate}) never changes sign this represents a deceleration.  However, for the case of a large outbound velocity, we would note an outward acceleration away from the central mass, rather than an attraction.  If the very high redshift on distant galaxies is interpreted as arising from a Doppler separation velocity, then the observed accelerating expansion of the universe, could be seen as an observer dependant affect at the Earth, given by Hilbert's equation.

Now, a more direct comparison with the Rindler frame would be given by a shell observer located at a fixed $ r $ in Schwarzschild coordinates at the location of the falling particle. We can write the line element as~\cite{mcgruder1982gravitational}
\be \label{ShellObserverMcGruder}
c^2 d \tau^2 = c^2 d T^2 - d R^2 - r^2 d \theta^2 - r^2 \sin^2 \theta d \phi^2 ,
\ee
where the conversion between the observer at a coordinate $ r $ and the observer at infinity is now $ d T^2 = \left (1-\frac{2 \mu}{r} \right ) d t^2 $ and $ d R^2 = \left (1-\frac{2 \mu}{r} \right )^{-1} d r^2 $.
Therefore the radial velocity is
\be \label{VelocityConversions}
\frac{d R}{d T} =  \left (1-\frac{2 \mu}{r} \right )^{-1} \frac{d r}{d t} = \frac{m c^2}{E_0} \frac{ d r}{d \tau }.
\ee
Therefore
\bea \label{AccelerationObserverAtr} \nonumber
\frac{d^2 R}{d T^2 } &= &  \frac{d R}{d T} \frac{d}{d R}\left (\frac{d R}{d T } \right ) = \frac{m^2 c^4}{E_0^2} \frac{d r}{d R} \frac{d}{d r}\left (\frac{d r}{d \tau } \right ) \frac{d r}{d \tau} \\ \nonumber
& = & -\frac{\mu c^2 }{r^2 }  \frac{m^2 c^4}{E_0^2} \left (1-\frac{2 \mu}{r} \right )^{\frac{1}{2}} \\
& = & -\frac{G M }{r^2 } \left (1-\frac{2 \mu}{r} \right )^{-\frac{1}{2}} \left ( 1 -\left (\frac{d R}{c d T} \right )^2 \right )  .
\eea
This last relation shows that for a local observer, the acceleration while apparently weakening at high velocity, never changes sign as for the case of Hilbert's equation for the observer at infinity.

This equation\footnote{Eq.~(\ref{AccelerationObserverAtr}) is a special case of the general result~\cite{mcgruder1982gravitational} restricted to purely radial motion.},
Eq.~(\ref{AccelerationObserverAtr}), is applicable to a terrestrial experiment located in the gravity field of the Earth measuring radially falling particles.
We can see, therefore, that the Schwarzschild solution gives velocity dependent geodesics for all observers at rest with respect to the gravitational field coordinates. This could be interpreted as an apparent weakening of the field strength in gravity, for objects with radial velocity.

We also show in Appendix~B, that for the general time independent metric, subject to the condition that they approach flat space for $ r \rightarrow \infty $, then we find this same velocity dependence, as in Eq.~(\ref{SchwarzschildVelocityDependenceWeakField}).
However, we may attempt to remove the velocity dependence of the acceleration by using the Painlev{\'e}-Gullstrand coordinate system, which matches coordinate time to proper time and is spatially flat, and given by
\be \label{gullstrandpainleve}
c^2 d \tau^2 = c^2 d t^2 - (dr - v dt)^2 ,
\ee
where $ t $ will be the time on the free-fall clock and $ v = -\sqrt{\frac{2 G M}{r}}  $ is defined as the escape velocity at each $ r $.  The transform between this metric and the Schwarzschild metric is given by 
\be \label{gullstrandpainleveGammaConversion}
dt \rightarrow dt +  \frac{v}{c^2 (1-v^2/c^2)} dr .
\ee 
It is interesting to note that the coefficient for $ dr $ differs from the derivative of gamma with respect to velocity,   by a factor of gamma, which would give us Eq.~(\ref{GravityAcceleration}), our original velocity dependence in flat space.  In these coordinates we find the coordinate acceleration now given by $ \frac{d^2 r}{d t^2} = -\frac{G m}{r^2} $ and so indeed now has no velocity dependence. 
However, we note that these coordinates assume a particle beginning at rest at $ r \rightarrow \infty $ and so more generally we will need to adjust the metric for particles with different input velocities to the field if we wish to remove the velocity dependence. This suggests that the velocity dependence has now moved to the metric itself.

In order to properly compare a Schwarzschild frame observer at some radius $ r_0 $ with the rocket
frame observer at $ X = 0 $ in flat space, we need to expand the Schwarzschild metric around $ r = r_0 $ and adjust
clock rates to measure proper time and rulers to measure proper distance~\cite{Moreau1994,Adler1991}. Thus, write the
Schwarzschild metric factor as
\bea
1 - \frac{2 G M}{c^2 r} & = & 1 - \frac{2 G M}{c^2 r_0 \left (1+\frac{x}{r_0} \right )} \\ \nonumber
& \approx & 1 - \frac{2 G M}{c^2 r_0}  \left ( 1 - \frac{x}{r_0} \right ) \\ \nonumber
& = & \left ( 1 - \frac{2 G M }{c^2 r_0} \right ) \left ( 1+ \frac{2 \overline{g} x/c^2}{1-\frac{2 G M}{c^2 r_0}} \right ) ,
\eea
where the approximate expression is valid if $ |x|/r_0 \ll 1 $ and we have defined $ \overline{g} = \frac{G M}{r_0^2} $.
The relationship between $ x $ and proper distance $ X $ is given by
\begin{equation}
X \approx \left ( 1 + \frac{G M}{c^2 r_0} \right ) x.
\end{equation}
Also letting
\begin{equation}
d T^2 = \left ( 1 - \frac{2 G M}{c^2 r_0} \right ) d t^2,
\end{equation}
finally gives a local transformed Schwarzschild metric of
\begin{equation} \label{SchwarzschildLocalExpansion}
c^2 d \tau^2 \approx \left (1+\frac{2 g X}{c^2} \right ) c^2 d T^2 - \frac{d X^2}{1+ \frac{2 g X}{c^2}} ,
\end{equation}
where $ g \approx G M/r_0^2 $
in the weak field limit. As expected, Eq.~(\ref{SchwarzschildLocalExpansion}) is Riemann flat~\cite{Adler1991,Moreau1994}.  We note, therefore, that a metric can be Riemann flat without a spatial metric coefficient of unity~\footnote{A suitable metric for a homogeneous gravitational field was
determined by Desloge~\cite{Desloge1987} as
\begin{eqnarray} \label{Equation11ReviewerTable}
c^2 d \tau^2 & = & \rme^{2 g X/c^2} c^2 d T^2 - d X^2, \\
 & \approx & \left (1+\frac{2 g X}{c^2} \right ) c^2 d T^2 - d X^2,
\end{eqnarray}
when $ |g X/c^2 | \ll 1 $.
Note, though, that it does not satisfy the vacuum Einstein equations. Also, in the weak field, we obtain the same velocity dependence as the Rindler metric and so this metric is not used.}
Incidentally, we can obtain this metric from the Schwarzschild metric in Eq.~(\ref{SchwarzschildMetric}), simply by making the substitution $ \frac{G M}{r} \rightarrow -g X $ describing a field of constant acceleration $ g $.

Now, making the coordinate transformation $ X \rightarrow X + \frac{c^2}{2 g} $ followed by $  X \rightarrow \frac{g}{2 c^2} X^2  $, in Eq.~(\ref{SchwarzschildLocalExpansion}), we obtain a form of the Rindler metric
\begin{equation} \label{RindlerStandardForm}
c^2 d \tau^2 = \frac{g^2 X^2}{c^2}  d T^2 - d X^2 .
\end{equation}
Alternatively, with a shift in the coordinate origin $ X \rightarrow X- c^2/g $ we find
\begin{equation} \label{RindlerStandardFormZeroed}
c^2 d \tau^2 = \left (1 +\frac{g X}{c^2} \right )^2 c^2 d T^2 - d X^2 ,
\end{equation}
produces another form of the Rindler metric.
Hence, we see that a local region of the Schwarzschild metric indeed matches the Rindler metric, confirming the equivalence principle.

Interestingly, even though we have applied a coordinate transform from the metric of a local region in the Schwarzschild metric in Eq.~(\ref{SchwarzschildLocalExpansion}) to the Rindler metric shown in Eq.~(\ref{RindlerStandardForm}) and Eq.~(\ref{RindlerStandardFormZeroed}), the calculations of the velocity dependence will give a factor of two as expected by Eq.~(\ref{AccelerationRindlerAction}) for Rindler but a factor of three for Schwarzschild metric shown in Eq.~(\ref{SchwarzschildVelocityDependenceWeakField}).  Note that even though Eq.~(\ref{SchwarzschildVelocityDependenceWeakField}) is a local Riemann flat region of the field we nevertheless produce the same velocity dependence of Eq.~(\ref{SchwarzschildVelocityDependenceWeakField}) and so different from the Rindler frame.
We can now see the cause of the discrepancy, that in converting from Schwarzschild metric in Eq.~(\ref{SchwarzschildLocalExpansion}) to the Rindler metric of Eq.~(\ref{RindlerStandardForm}) we applied a quadratic coordinate transformation of $  X \rightarrow \frac{g}{2 c^2} X^2  $ and so modified the velocity dependence of coordinate acceleration.  Hence the variation in the velocity dependence of the coordinate acceleration indeed appears to be a spatial coordinate effect rather than spatial curvature.  This also then explains why when we set the spatial coefficient to one in the Schwarzschild metric, as shown in Appendix~A and Eq.~(\ref{SchwarzschildVelocityDependenceAppendixFlatSpace}), we find the velocity dependence becomes $ 1 - \frac{2 v^2}{c^2}  $ in the weak field, and so now matching the result for the Rindler frame. 

%The last conundrum is that the shell observer for the  Schwarzschild solution, shown in Eq.~(\ref{AccelerationObserverAtr}) of %$ 1 - \frac{ v^2}{c^2}  $ gives a factor of one for the velocity dependence.  

\subsection{Non curved solutions to uniform gravitational fields and exact equivalence to accelerated frames}

A key expectation of this paper is that there should be an analogous result in gravity for the velocity dependence of acceleration in special relativity. This approach, as we have seen above is not necessarily easily implemented. One of the complicating factors is the insistence by some authors, such as  Desloge, who  disallow a flat space to exist in a uniform gravitational field. On the other hand Munoz and Jones and others~\cite{munoz2010equivalence,tilbrook1997general,rohrlich1963principle}
address Deloge's claims,
 finding solutions ~\cite{desloge1989nonequivalence} that produce a flat space in uniform gravitational fields consistent with general relativity and the Einstein vacuum solutions $ R_{\mu \nu}=0  $.
 
After considering the principles of isotropy, a static metric, translational invariance along planar spatial cross sections of the geometry, and orthogonal planes to the $x$-coordinate implying the metric coefficients depend only on $x$, Munoz and Jones begin with the metric
\begin{equation} \label{MunozMetric}
ds^{2}=\lambda^{2}(x)dt^{2}-\sigma^{2}(x) dx^{2}-\xi^{2}(x)(dy^{2}+dz^{2}) .
\end{equation}
Then, after using Taub’s theorem~\cite{taub1951empty}, which states that a spacetime with plane symmetry with  $R_{\mu \nu}=0  $ admits a coordinate
system where the line element is static, derives the following metrics which are equivalent to the Rohrlich-Tilbrook form~\cite{munoz2010equivalence}
\begin{equation} \label{RohrlichMetric}
ds^{2}=\lambda^{2}(x)dt^{2}-\sigma^{2}(x)dx^{2}-dy^{2}-dz^{2} ,
\end{equation}
with $ \lambda'=g\sigma $.

Equation  Eq.~(\ref{RohrlichMetric}) is the only admissible one representing a singularity free, flat spacetime of a uniform gravitational field, consistent with general relativity.

Unlike the Rindler metric, given by Eq.~(\ref{RindlerStandardForm}), these equations do satisfy the  Einstein vacuum solutions.
Furthermore, after considering a free falling observer in a gravitational field, it is possible to derive the fundamental time asymmetric acceleration equations that produces the usual hyperbolic motions of accelerated systems
as observed by inertial observers in flat space $ T=\frac{\lambda(x)}{g} \sinh(gt)$.
With this elegant result, and since the form of these equations can be recovered from Eq.~(\ref{SRAcceleration})  by a second integral~\footnote{We have $a=g(1-\frac{v^{2}}{c^{2}})^{\nicefrac{3}{2}}=\nicefrac{dv}{dt}  $ and so
$ g dt=dv(1-\frac{v^{2}}{c^{2}})^{-\nicefrac{3}{2}}$ and so integrating gives $gt=\frac{v}{\sqrt{1-\frac{v^{2}}{c^{2}}}} $. Hence $T= \intop \frac{dt}{\sqrt{1-\frac{v^2}{c^2}}}=\intop\frac{dt}{\sqrt{1+(\frac{gt}{c})^2}}=\frac{c}{g} \sinh^{-1}\left (\frac{gt}{c} \right )$ .}, one is tempted to place the primary cause of the velocity dependence of a point particle, on the asymmetry in the time coordinates.  Additionally this idea also confirmed in the last section, where we found going from Eq.~(\ref{SchwarzschildLocalExpansion}) to  Eq.~(\ref{RindlerStandardForm}) involved a spatial coordinate effect, rather than spatial curvature.  
However our other proposition that the velocity dependence is in the metric itself, further supported by  the Munoz and Jones analysis, suggests that the additional curvatures of the metric, e.g. Schwarzschild, adds to the velocity dependence, `on top' of the coordinate contribution. In view of this we would expect the results to match very closely for approximate uniform sections of the gravitational field and accelerations frames. We therefore  propose the effect would make an interesting test of general relativity, testable under Earth's gravity.

\section{Discussion}

We show, in this paper, using the principle of equivalence to relate accelerating frames and gravity, the velocity dependence of coordinate acceleration under gravity, for weak locally homogeneous regions of gravitational fields. Specifically, we have shown that the behavior of inertial free fall particles in gravity are a function of initial particle velocity, as given by Eq.~(\ref{GravityAcceleration}). We also note the comparison to a little-known result first derived by Hilbert in 1917, shown in Eq.~(\ref{SchwarzschildVelocityDependenceWeakField}). From the viewpoint of a shell observer, fixed with respect to Schwarzschild coordinates, this might also be interpreted as a weakening of the field. The discrepancy factor of two for velocity dependence of acceleration in the Rindler frame Eq.~(\ref{AccelerationRindlerAction}) of $ 1 - \frac{2 v^2}{c^2 } $ compared with the equivalent shell observer for Schwarzschild in Eq.~(\ref{AccelerationObserverAtr}) of $ 1 - \frac{ v^2}{c^2} $, shows the difference in physical effects that a curved and flat space can manifest. This also points to the limits of applicability when attempting to apply the principle of equivalence.

We might expect the velocity dependence of acceleration, due to coordinate acceleration for timelike radial geodesic motion, can be obtained from the Christoffel symbols by contraction with two velocity vectors within the geodesic equation. Nevertheless, the acceleration in terms of proper time is the Newtonian relation in Eq.~(\ref{SecondDerivativeProperTimeMain}), which contains no velocity dependence.  The velocity dependence only appears when we convert to a coordinate time, as shown in Eq.~(\ref{SchwarzschildVelocityDependenceAppendix}).  Also, the relation of velocity dependence as found in a Rindler frame in special relativity in flat space, is comparable to that found under the Schwarzschild solution of general relativity.  This appears to indicate that the source of the velocity dependence can be identified within special relativity, and specifically appears to relate to the use of a coordinate time parameterization of the path as opposed to proper time. Also, the fact that an exact correspondence is found between a Rindler frame in flat space and general relativity for arbitrary initial velocities, shows the generality of the equivalence principle as applying to not just particles falling from rest, but also for arbitrary initial radial velocities.
Interestingly, we can remove the velocity dependence by changing to the Painlev{\'e}-Gullstrand coordinate system, although the velocity dependence is now in the metric itself. We could also see this in flat space where the velocity dependence can be viewed as a boost out of the comoving frame. 
We have also shown in Appendix~B that independent of coordinate choices in the Schwarzschild metric, we recover in the weak field, the same velocity dependence as found in Eq.~(\ref{SchwarzschildVelocityDependenceWeakField}). We also identified the variation in the velocity dependence of coordinate acceleration, shown by Eq.~(\ref{AccelerationRindlerAction}) for the Rindler frame and Eq.~(\ref{SchwarzschildVelocityDependenceWeakField}) for the Schwarzschild observers, as due to the spatial coordinate transformation between the frames and not spatial curvature {\it per se}. Given that flat space solutions do exist in uniform gravitational that are in exact correspondence with acceleration where we know velocity dependence exists, we see that Eq.~(\ref{SRAcceleration}) and Eq.~(\ref{SchwarzschildVelocityDependenceAppendixFlatSpace}) should hold in such regions of gravity. This assists in understanding some of the discrepancies between several of our results, such as the coordinate dependencies for flat space versus spatial metrical effects when curvature is taken into account using the Schwarzschild metric. Additionally we also show our approach aids in giving insight into how geodesic deviation within general relativity might behave in Appendix~C. 

Regarding experimental tests, we derive Eq.~(\ref{AccelerationObserverAtr}), which gives the acceleration expected for a terrestrial experiment for the velocity dependence of acceleration due to gravity.
Currently, lunar and satellite laser ranging, applied to the radial component of orbital motion, allows precise verification of GR and the EEP using the parameterized post Newtonian or PPN formalism.  However, the velocity dependence effect could be explicitly tested, such as by measuring free-fall in a vacuum with high  precision  sapphire clocks now accurate to 1 part in $10^{-18}$ seconds~\cite{takamizawa2014atomic}. This type of test would also provide further confirmation of the EEP.

We believe our approach provides a natural pathway from accelerations in special relativity to the behavior of radial geodesics in gravitational fields, including tidal effects.  Of course, when making these comparisons, since ruler lengths and clock rates can differ between the two frames, we need to be careful that we are identifying physical agreement rather that merely formal agreement.

\vskip 1pc

%\ethics{Does not apply to this manuscript.}

%\dataccess{``This article has no additional %data''.}

%\aucontribute{DLB developed original concept, %JMC produced 
%initial draft paper,
%adding further results, and DA supervised the %study.  All authors 
%checked and proofed the
%manuscript.}

%\competing{No competing interests.}

%\funding{None.}
%\ack

\pagebreak

\appendix

\section{Geodesic motion using a Lagrangian approach}

Geodesics follow paths of maximal proper time, and from the line element in Eq.~(\ref{SchwarzschildMetric}) we produce a Lagrangian
\be \label{RearrangedMetricLagrangian}
\mathcal{L} = -g_{\mu \nu} \dot{x}^{\mu} \dot{x}^{\nu} = \left (1-\frac{2 \mu}{r} \right ) \dot{t}^2 - \frac{1}{c^2} \left (1-\frac{2 \mu}{r} \right)^{-1} \dot{r}^2 = 1 ,
\ee
where $ \dot{t} = \frac{d t}{d \tau} $ and $ \dot{r} =  \frac{d r}{d \tau} $ and for purely radial motion we have assumed that the angular terms are zero.  As we are assuming we are dealing with particles with mass we can parameterize their motion using the proper time $ \tau $ and so we can use $ \mathcal{L} = -g_{\mu \nu} \dot{x}^{\mu} \dot{x}^{\nu} $ rather than $ \mathcal{L} = \sqrt{-g_{\mu \nu} \dot{x}^{\mu} \dot{x}^{\nu} } $.

Lagrange's equations are an equivalent form of the geodesic equation Eq.~(\ref{GeodesicEquation}), and for $ t $ produce $ \frac{d}{d \tau} \left ( \frac{\partial \mathcal{L}}{\partial \dot{t} } \right ) - \frac{\partial \mathcal{L}}{\partial t} = 0 $, giving
\be
\frac{d }{d \tau} \left ( \left (1-\frac{2 \mu}{r} \right )c^2 \dot{t} \right ) = \frac{d \mathcal{L}}{d t} = 0 .
\ee
Hence we have a constant of the motion 
\be \label{ConstantOfMotion}
\left (1-\frac{2 \mu}{r} \right ) \dot{t} = \frac{E_0}{m c^2} ,
\ee
where $ E_0 $ represents the conserved total energy for its motion.
Substituting Eq.~(\ref{ConstantOfMotion}) into Eq.~(\ref{RearrangedMetricLagrangian}) we find
\be \label{VelocityProper}
\frac{ d r}{ d \tau} = \pm c \sqrt{\frac{E_0^2}{m^2 c^4} -  \left (1 - \frac{2 \mu}{r} \right ) } .
\ee
Differentiating with respect to proper time $ \tau $ gives
\be
a^r = \frac{ d^2 r }{d \tau^2 } =  \frac{ d r}{ d \tau}  \frac{ d }{d r } \left (\frac{ d r}{ d \tau} \right )= -\frac{\mu c^2 }{r^2 } ,
\ee
as measured by a co-moving observer.

Now, using Eq.~(\ref{ConstantOfMotion})  we find
\be \label{Appendixdrbydt}
\frac{d r}{ d t} =  \frac{d \tau}{ d t} \frac{d r}{ d \tau} =  \pm c \left ( 1 - \frac{ 2 \mu }{ r} \right ) \sqrt{ 1 - \frac{m^2 c^4}{ E_0^2 } \left (1 - \frac{2 \mu}{r} \right ) } .
\ee

Then, using the chain rule $ \frac{d^2 r }{d t^2 } = \frac{d (dr/dt) }{d r } \frac{ d r }{ d t} $, we find
\be \label{SchwarzschildCoordinateAccelerationAppendix}
\frac{d^2 r}{d t^2} = - \frac{\mu c^2}{r^2 } \left ( 1 - \frac{2 \mu}{r} \right ) \left ( 3 \left ( 1 - \frac{2 \mu}{r} \right )\frac{m^2 c^4}{E_0^2} -2 \right ) .
\ee
Finally, using the relation between $ E_0 $ and $ \frac{dr}{dt} $ in Eq.~(\ref{Appendixdrbydt}) we find
\be \label{SchwarzschildVelocityDependence}
\frac{d^2 r}{d t^2} = - \frac{\mu c^2}{r^2 } \left ( 1 - \frac{ 2 \mu}{r} \right ) \left (1 - \frac{3 }{c^2} \left (\frac{ d r}{d t} \right )^2 \left ( 1 - \frac{ 2 \mu}{r} \right )^{-2} \right ) ,
\ee
in agreement with Eq.~(\ref{SchwarzschildVelocityDependenceAppendix}).

If we set $ g_{rr} = -1 $ and repeat the derivation above, then we find $ \frac{d t}{d \tau } $ is unaffected but we now have
\bea 
\frac{d r}{d \tau } & = &  \pm c  \left (1 - \frac{2 \mu }{r} \right )^{-1/2} \sqrt{ \frac{ E_0^2}{ m^2 c^4 } - \left (1 - \frac{2 \mu }{r} \right ) } \\ \nonumber
\frac{d^2 r}{d \tau^2 } & = &  - \frac{G M}{r^2} \frac{ E_0^2}{ m^2 c^4 }  \left (1 - \frac{2 \mu }{r} \right )^{-2} \nonumber
\eea
and
\be \label{SchwarzschildVelocityDependenceAppendixFlatSpace}
\frac{d^2 r}{d t^2} = - \frac{\mu c^2}{r^2 } \left (1 - 2  \left (\frac{ d r}{ c d t} \right )^2 \left ( 1 - \frac{ 2 \mu}{r} \right )^{-1} \right ) .
\ee
Hence in the weak field limit, the coefficient on the velocity dependence reduces from 3 to 2 in this case, now agreeing with the flat space Rindler frame in Eq.~(\ref{AccelerationRindlerAction}).  We note, though, that a spatially flat space can be achieved with $ g_{rr} \ne 1 $.

\section{General form of the Schwarzschild metric}

The Schwarzschild metric in its most general form, which is the most general spherically symmetric vacuum solution of the Einstein field equations for a non-rotating uncharged mass, can be written
\bea   \label{SchwarzschildMetricGeneral}
  c^2 d\tau^2 & = & \left (1-\frac{2 \mu}{D} \right )c^2dt^2 - {D'}^2 \left (1-\frac{2 \mu}{D} \right)^{-1} dr^2 \\ \nonumber
	& & - D^2 d\theta^2 -    D^2\cos^2\theta d\phi^2. \nonumber
\eea
The four common variants, which are time independent, are:  Schwarzschild's original metric with $ D = (r^3 + 8 \mu^3)^{1/3} $, isotropic coordinates with $ D = r (1 + \frac{\mu}{2 r})^2 $, Brillouin coordinates with $ D = r + 2 \mu $ and the more conventional form of the metric with $ D  = r $.  These coordinate choices all enforce a flat space condition, such that for  $ r \rightarrow \infty $, $ D \rightarrow r $.  This  also implies that $ D' \rightarrow  1 $ and $ D'' \rightarrow 0 $ as $ r \rightarrow \infty $. 

Now, given alternative spatial coordinates $ \rho = D(r) $, we have $ d \rho = D' d r $, where $ D' = \frac{d D}{d r}  = \frac{d \rho}{d r} $, then
\be 
\frac{d^2 \rho}{dt^2} = \frac{ d}{d t} \left (\frac{ dr }{dt} \frac{d \rho}{ d r} \right ) = \frac{d \rho}{ d r} \frac{d^2r}{dt^2} + \left (\frac{dr}{dt} \right )^2  \frac{ d}{d r} \left ( \frac{d \rho}{ d r}\right ) .
\ee 
Hence, in the weak field with $ \frac{ d \rho}{ d r}  = $ constant the second term is zero and so we recover the same velocity dependence, as found in Eq.~(\ref{SchwarzschildVelocityDependenceWeakField}).
This implies we limit new coordinate systems, in the weak field regime far from the source, to $ \rho = a + b r $, where $ a , b $ are constants, giving $ \frac{ d \rho}{ d r} = b = $ constant, as required.
Hence, we have shown, independent of coordinate choices in the Schwarzschild metric we have recovered, in the weak field, the same velocity dependence as found in Eq.~(\ref{SchwarzschildVelocityDependenceWeakField}).

The general Christoffel symbols are:
\be
\Gamma^r_{rr} = \frac{D''}{D'} - \frac{\mu D'}{ D(D-2 \mu)} \, ,\,
\Gamma^r_{tt} =   \frac{c^2  \mu (D-2 \mu)}{ D' D^3} \, ,\,
\Gamma^r_{rt} =   \frac{ \mu D'}{ D(D-2 \mu)}  .
\ee
This gives the radial acceleration
\be
\frac{d^2 r}{d \tau^2} = - \left (\frac{c^2  \mu (1-2 \mu/D)}{ D' D^2} \right )\left(\frac{d t}{d \tau} \right )^2 - \left (\frac{D''}{D'} - \frac{\mu D'}{ D^2(1-2 \mu/D)} \right )\left(\frac{d r}{d \tau} \right )^2 .
\ee
This then recovers the Newtonian acceleration $ \frac{d^2 r}{d \tau^2} = -\frac{G M}{r^2} $ in the weak field limit independent of $ D $.

\section{Velocity dependence of tidal force} 

Up to this point, apart from Eq.~(\ref{SRAccelerationInvariant1}), we have only dealt with coordinate dependent effects. However in gravity there is a frame in which the effect cannot be transformed away by coordinate boosts. These are the frames that experience tidal accelerations, which in General Relativity characterize gravity and the curvature of spacetime, via geodesic deviation. We give only an elementary investigation of such effects here, with the aim of showing that under gravity, the individual coordinate dependent effects do affect coordinate independent tidal accelerations.This section is also based on the work of authors such as Munoz that show the gravitational field has solutions consistent with uniform acceleration.

Following on from section \ref{thoughtExperiment} we consider a group of PG1$_n$ evenly separated, and each occupying a point in flat spacetime. Together these form a simple inertial frame in which all clocks can be synchronized and agree on simultaneity. Therefore if a point is accelerated past this frame, then the accelerated point would encounter PG1$_1$, PG1$_2$.....PG1$_n$. Hence if PG1$_n$ measures the acceleration from Eq~(\ref{GravityAcceleration}), then all clocks in this inertial frame can agree exactly when the particle passed the point PG1$_n$. If, for example, the accelerating point passes PG1$_a$ before PG$_b$ then PG1$_b$ will measure a slightly smaller acceleration for the point on the rocket compared to PG1$_a$. Now, invoking the principle of equivalence for each individual particle, the PG1$_n$ individually follow geodesics that when close together form an approximate inertial reference frame consistent with the laws of special relativity. Therefore to a good approximation the PG1$_n$ are still able to agree on their respective clock synchronizations, and hence on observations they make of their respective observed accelerations and individually they will observe acceleration dependence on velocity.

Since the PG1$_a$, PG1$_b$ are at slightly different positions in the field and therefore at slightly different potentials, they will accelerate at slightly different rates, and hence experience geodesic deviation. Therefore while individually the PG1$_n$ are coordinate dependent, together they are not. This leads to a natural investigation of how the velocity dependencies of each individual PG1$_n$, affect the geodesic deviation between them.

The Newtonian equation for tidal accelerations, based on the gravitational acceleration, around a mass $ M $, of $ a = -\frac{G M}{r^2 } $, is \be \label{NewtonianTidal} \frac{d a}{d r} = \frac{2 G M}{r^3 } . \ee This implies that for objects at a radial separation $ dr $ in a gravitational field experience a differential acceleration $ d a $. 

In the domain of special relativity, we can treat Eq.(\ref{GravityAcceleration}) as a relativistic correction to an acceleration akin to Newtonian gravity, implying a tidal acceleration 
\be \label{FinalDaveTidalSolution} \frac{d a }{d r} = \frac{ 2 G M }{\gamma^3 r^3 } \left [1- \frac{3}{4 \gamma}  \frac{2 G M}{r c^2} \right ] \approx \frac{ 2 G M }{\gamma^3 r^3 }, 
\ee 
where we have neglected the effect of the second term in brackets.  This is justified as in the weak field we can assume $ r \gg \frac{2 G M}{c^2} = r_s $, where $ r_s $ is the Schwarzschild radius and also for relativistic velocities $ \frac{1}{\gamma} \rightarrow 0 $. Therefore, special relativity adds to the Newtonian result a velocity dependent term, which implies that tidal forces will appear to reduce for higher velocities. Also of interest is that the Schwarzschild radius $ r_s $ arises naturally within special relativity with a Newtonian acceleration profile.

In the low velocity limit weak field limit, this approximates to
\be \label{FinalDaveTidalSolutionLowVelocity} \frac{d a }{d r} \approx \frac{ 2 G M }{ r^3 } \left ( 1 - \frac{3 v^2 }{2 c^2 } \right ) , \ee 
showing the velocity dependence compared with the Newtonian result in Eq.~(\ref{NewtonianTidal}).

As we have solved the geodesic equation in Eq.(\ref{SchwarzschildCoordinateAcceleration}), which gives the one-dimensional radial acceleration as a function of $ r $, we can find the relative acceleration with respect to any fixed point in the field by differentiation with respect to the radius $ r $, giving 
\be \frac{ d a}{ d r}= \frac{2 G M }{r^3} \left ( \frac{3 m^2 c^4 }{E_0^2} \left (1- \frac{ r_s}{r} \right ) \left (1- \frac{2 r_s}{r} \right ) - \left (2-\frac{3 r_s}{r} \right ) \right ) . \ee 
Substituting for the total particle energy $ E_0 $ in terms of velocity in Eq.~(\ref{VelocityCoordinate}), we find the effective tidal acceleration \be \label{TidalAccelerationGRV} \frac{d a}{d r} = \frac{2 G M }{r^3} \left ( 1 - \frac{3 r_s}{r} - \frac{3 v^2}{c^2 } \frac{ \left ( 1 - \frac{2 r_s}{r} \right ) }{ \left (1 - \frac{r_s}{r} \right )^2} \right ) , 
\ee 
where $ v = \frac{d r}{d t} $.

For the weak field case where $ r \gg r_s $, we have the effective tidal acceleration 
\be \label{WeakFieldTidalGR}
\frac{d a}{d r} \approx \frac{2 G M }{ r^3} \left ( 1 - \frac{3 v^2 }{c^2} \right ). 
\ee 
Hence the velocity dependence in Schwarzschild coordinates, in the weak field limit, is twice that obtained previously using special relativity as shown in Eq.~(\ref{FinalDaveTidalSolutionLowVelocity}). Hence, the special relativistic approach gives qualitatively the same result as general relativity, with the commonly recurring factor of two discrepancy. This is therefore an alternative approach to gain insight into geodesic deviation, based on velocity dependence within special relativity.

\bibliographystyle{RS}
\bibliography{quantum}

%\begin{thebibliography}{99}

%\bibitem{Schutz} B.~F.~Schutz, {\it A first course in general
%  relativity} (Cambridge University Press, Cambridge, 1988)

%\end{thebibliography}

\end{document}